\documentclass[aps, twocolumn, superscriptaddress, pre, reprint, floatfix, numbers, sort, compress]{revtex4-2}


\usepackage{graphicx}
\usepackage{amssymb,amsmath,amsbsy,amsfonts,bm}
\usepackage[colorlinks=true,linkcolor=black,citecolor=black,urlcolor=black]{hyperref}
\usepackage{epsfig}
\usepackage{multirow}
\usepackage{tikz}
\usetikzlibrary{calc}
\usepackage{threeparttable}
\usepackage{adjustbox}
\usepackage{booktabs}
\usepackage{diagbox}
\usepackage{array}
\usepackage{lipsum}
\usepackage[normalem]{ulem}

\usepackage[percent]{overpic}
\usepackage{color}

\newcommand{\totdiff}[2]{\frac{\text{d }#2}{\text{d }#1}}
\newcommand{\MSD}{\text{MSD}}
\newcommand{\MSJ}{\text{MSJ}}
\newcommand{\Wkin}{W_\text{kin}}
\newcommand{\Tkin}{T_\text{kin}}

\begin{document}
\title{Gigantic dynamical spreading and anomalous diffusion of jerky active particles}

\author{Hartmut L\"owen}
\affiliation{Institut f\"ur Theoretische Physik II: Weiche Materie, Heinrich-Heine-Universit\"at D\"usseldorf, Universit\"atsstra{\ss}e 1, 
40225 D\"usseldorf, Germany}

\begin{abstract}
Jerky active particles are Brownian self-propelled particles which are dominated by ``jerk'', the change in acceleration. They represent a generalization of inertial active particles. In order to describe jerky active particles, a linear jerk equation of motion which involves a third-order derivative in time, Stokes friction and a spring force is combined with activity modeled by an active Ornstein-Uhlenbeck process. This equation of motion is solved analytically and the associated mean-square displacement (MSD) 
is extracted as a function of time. For small damping and small spring constants, the MSD shows an enormous superballistic spreading with different scaling regimes characterized by anomalous high dynamical exponents 6, 5, 4 or 3 arising from a competition between jerk, inertia and activity. 
When exposed to a harmonic potential, the gigantic spreading tendency induced by jerk gives rise to an enormous increase of the kinetic temperature and even to a sharp localization-delocalization transition, i.e. a jerky particle can escape from harmonic confinement.  The transition can be either first or second order as a function of jerkiness. Finally it is shown that self-propelled jerky particles governed by the basic equation of motion can be realized experimentally both in feedback-controlled macroscopic particles and in active colloids governed by friction with memory.
\end{abstract} 

\maketitle

\section{introduction}
One of the immediate consequences of Newton’s postulates \cite{Newton} in classical mechanics implies that a force field can only depend on the position and velocity of the particle but not on its acceleration. Assuming such an acceleration-dependence would immediately be in conflict with the superposition principle \cite{Newton,Rebhan,acceleration}. Another basic assumption is that the inertial mass of the particle is positive. Since Newton’s axioms, however, only hold for the fundamental interactions, in general this can be challenged in artificial or effective force fields. Likewise, despite Newton’s third law ``actio=reactio'', there are many current studies on nonreciprocal interactions which violate Newton’s third laws, such as predator-prey systems. This is not a fundamental violation but occurs in effective coarse-grained nonequilibrium situations. Still this requires a re-thinking about the foundations of mechanics and statistical mechanics in nonequilibrium and many novel effects have been discovered due to non-reciprocity \cite{Ivlev,Tailleur,Stark,Loos,Golestanian,Kardar,roadmap2}.

Here we challenge the two basic assumptions of acceleration-independent forces and positivity of inertial mass. These challenges can be addressed using programmable active robots subjected to a time-delayed feedback force specifically designed to break the assumption of acceleration-independent force fields. This is simulating an effective inertial mass with any sign, i.e. it can even realize an equation of motion for a particle which is formally identical to Newton’s equation of motion but with a negative particle mass. 
Such a negative-mass particle is found within the context of external feedback control. By monitoring the trajectories of the particle externally with a camera, one can measure its acceleration and compute the corresponding force. Then the particle is exposed to an external field that is exactly reflecting the computed force \cite{Bechhoefer,feedback,Babel,Komura_2024,Komura_2025}. These artificial forces then formally break the condition that forces are acceleration-independent and that mass is positive. When combined with a short time delay, these synthetic forces involve the {\bf jerk} which is the time derivative of the acceleration \cite{Schot_1978,Linz_1998} resulting in equations of motion that include the third-order time derivative \cite{Linz_1997,Eichhorn_1998,Eichhorn_2002,Umut,Pham,jerk_10,jerk_11,jerk_12,jerk_13} rather than the second order as is typical in Newtonian mechanics. 
So far, jerky particle dynamics have been primarily investigated within the framework of chaos theory in nonlinear dynamical systems \cite{Linz_2000,Pramana,chaos1,chaos2}, traffic flow control \cite{traffic1,traffic2}, the Abraham–Lorentz force associated with radiation emission \cite{traffic1,traffic2}, and cosmology \cite{cosmology1,cosmology2}.

Another rapidly growing research field is the physics of active matter. Active matter consists of active particles that convert energy from the environment to mechanical motion, as for reviews see, e.g.,  \cite{Ramaswamy,Marchetti,Elgeti,Bechinger_RMP_2016,roadmap,Vrugt_review}. Modelling of active particles can occur on various levels, maybe the simplest model is the active Ornstein-Uhlenbeck particle which is based on memory-dominated noise studied for passive particles already by Fürth in 1920 \cite{Fuerth} and by Ornstein and Uhlenbeck in 1930 \cite{Ornstein}. This model describes a particle that is randomly exposed to persistent noise providing kicks into the same direction during a typical time scale, the so-called {\it persistence time} $\tau_p$.
The active Ornstein-Uhlenbeck model has been applied to many situation in order to describe activity in a simple way, see e.g.  \cite{Szamel,Bonilla,Fodor,Dabelow,Komura_2024,Wittmann,Seifert,Gupta_Klapp,Crisanti,Metzler3}.

 Here we combine the two fields of jerky dynamics and active particles. In this spirit, the term {\it jerky active matter} was coined and introduced by te Vrugt, Jeggle and Wittkowski  \cite{teVrugt_2021}. While they focused more on field theoretical equations describing active matter with an additional time scale, our work predominately investigates the individual particles and their stochastic active motion. We therefore consider a new model class of active particles which is dominated by jerk, and thus consists of {\it active jerky particles}.   
 
In doing so, we study the simplest model conceivable for activity namely an active Ornstein-Uhlenbeck particle subject not only to inertia \cite{Caprini_JCP_inertial,Nguyen,Lowen_review,Sandoval1,Sandoval2} but also to jerk. An appropriate linearized equation of motion  \cite{noisy_jerk_oscillator}  with activity is proposed and solved. One characteristic dynamical quantity to classify active particle dynamics is their mean-square displacement (MSD). Typically it involves different scaling exponents as a function of time $t$. Characteristic for a Brownian active particle is a crossover from a ballistic regime where the MSD scales with $t^2$ to a diffusive regime with a scaling linear in time $t$ \cite{Howse,Bechinger_RMP_2016}. With a term linear in the jerk in the equation of motion, the MSD shows an {\it enormous spreading} revealed by a scaling in time with unusual high exponents 6, 5 or 3 depending on whether there is pure jerk or a combination of jerk and inertia. These unusual exponents occur even in the long-time limit and characterize {\it anomalous diffusion} \cite{Metzler1,Metzler2,Babel_2014}. This high diffusion and corresponding spreading tendency is gigantic as it can even break the localization of a harmonically bound particle leading to a localization-delocalization transition for increasing jerk strength. Several experimental realizations of jerky active particles described by the  dynamics are also proposed.

The paper is organized as follows. In section II we present the model and introduce the mean-square displacement, the mean kinetic energy and the mean-square jerk. Different special cases are subsequently discussed and solved in section III. Section IV treats the general case of a jerky inertial damped harmonic oscillator and conclusions are given in section V.


\section{Active jerky particle dynamics}
\subsection{Equations of motion}
We consider a particle in one spatial dimension with a time-dependent position coordinate $z(t)$. Suppose an acceleration-dependent force is programmed externally by a feedback device which measures the instantaneous particle acceleration ${\ddot z}(t)$ by monitoring the change in the particle translational velocity ${\dot z}(t)$. This artificial force violates the traditional view of classical Newton’s mechanics where the force field does not depend on acceleration. Within a small time delay $\delta t >0$ the device transforms this input into a force $F( {\ddot{z}}(t-\delta t) )$ that is acting on the particle. For small deviations around a reference acceleration ${\ddot z}_0$ we can double Taylor-expand this function in acceleration and time to obtain the linear expression
\begin{equation}
    \label{eq:E1}
    F( {\ddot{z}}(t-\delta t) ) \approx F({\ddot z}_0) + F^{\prime}({\ddot z}_0) ({\ddot z}(t) -\delta t {\dddot z}(t)).
\end{equation}
where $(...)^{\prime}$ denotes a derivative with respect to acceleration. Now we superimpose this force with a Stokes friction force $-\gamma {\dot z}(t)$ where $\gamma$ is an appropriate friction coefficient and a harmonic spring force $-k {z}(t)$ with $k$ denoting the spring constant. 
The full stochastic equation of motion is obtained as a force balance
\begin{equation}
    \label{eq:(X0)}
    \lambda {\dddot x}(t) + m {\ddot x}(t) + \gamma{\dot x}(t) + k {x}(t) = \gamma  u(t)
\end{equation}
with a new shifted position coordinate $x(t)=z(t) - (F({\ddot z}_0)- F^{\prime}({\ddot z}_0){\ddot z} )/k$, a renormalized inertial mass 
\begin{equation}
    \label{eq:(X2)}
    m=m_p - F^{\prime}({\ddot z}_0)
\end{equation}
where $m_p$ is the real mass and the jerk coefficient
\begin{equation}
    \label{eq:(X1)}
    \lambda = F^{\prime}({\ddot z}_0) \delta t .
\end{equation}
 Two remarks are in order: first of all, the presence of the term $\lambda {\dddot x}(t)$ involves the ``jerk'', i.e. the change in acceleration and therefore defines the novel class of jerky particles. Thus the equation of motion goes beyond that of the well-known noisy harmonic oscillator \cite{colored,Fangfu}. A similar jerk dynamics with more generalized noise but nonlocal friction was studied recently in Ref.\ \cite{noisy_jerk_oscillator}.   Second, Eqn.~\eqref{eq:(X2)} implies that the {\it effective inertial mass can be negative\/}  provided $F^{\prime}({\ddot x}_0)$ exceeds the bare inertial mass. This is atypical for Newtonian dynamics. In principle both signs of $m$ and $\lambda$ can be steered independently at wish by the feedback function. In particular, for very small delay times $\delta t$ the inertial term can even dominate the jerk-term such that quick acceleration feedback can be used to steer a pure {\it negative inertial particle mass\/} without any jerk which is interesting in itself. 
 
The right-hand-side $\gamma u(t)$ of \eqref{eq:(X0)} is an {\bf activity\/} force (sometimes called swim force) which brings the particle to a self-propulsion velocity $u(t)$. The stochastic self-propulsion velocity $u(t)$ is described by an active Ornstein-Uhlenbeck process (AOUP) \cite{Fuerth,Ornstein,Caprini_JCP_inertial,Nguyen} as
\begin{equation}
    \label{eq:(AOUP)}
    \tau_p{\dot u}(t) = -{u}(t) + \zeta(t).
\end{equation}
Here $\tau_p$ is the intrinsic  persistence time and $\zeta(t)$ is a Gaussian white noise with zero mean and variance $\overline{\zeta(t)\zeta(t')} = 2v_0^2\tau_p \delta (t-t') = 2D \delta (t-t')$ where the overbar denotes a stochastic average over the noise and $D=v_0^2\tau_p$ is an effective diffusion coefficient. The parameter $v_0$ denotes a typical self-propulsion velocity of the active particle. Alternatively one can say that $u(t)$ represents colored noise \cite{Haenggi} since \cite{Nguyen}
\begin{equation}
    \overline {u(t) u(t')} =  v_0^2 \ \exp ( - | t  -  t'| /\tau_p).
\end{equation}
We denote a noise memory kernel of the activity by
\begin{equation}
    \label{eq:(M1)}
    M(t) = \gamma^2  v_0^2 \exp ( - |t| /\tau_p).
\end{equation}
If the persistence time is very large, much larger than any other time scale of the system, we call the particle ``persistent''. In this limit  of {\it persistent noise} the activity force memory kernel is constant
\begin{equation}
    \label{eq:(M2)}
    M_p(t) \equiv M(0) = \gamma^2 v_0^2 .
\end{equation}
  In the opposite limit of a very small persistence time $\tau_p$, taken such that $D=v_0^2\tau_p$ stays constant, we obtain (Markovian) {\it white noise} characterized by the force memory kernel 
\begin{equation}
    \label{eq:(M3)}
    M_w(t) = 2\gamma^2 D  \delta(t).
\end{equation}
We refer to this noise as {\it passive noise} and correspondingly the particle will be called ``passive''. 
This special case was treated previously with jerk-like dynamics in Ref.\ \cite{Caprini_flocking}. Moreover the action of deterministic forces in jerk-like equations were considered earlier, see e.g.\ \cite{Matej}. 

Note that far from equilibrium there is no fluctuation-dissipation relation. In general, the noise strength $\gamma v_0$ is not related to any friction coefficient and can even be steered artificially by external fields (see \cite{Argun,Isa1,Isa2} for example for such an experimental realization) or by external vibrations \cite{Scholz,Davide_vibration}. This makes our work different to previous modelings \cite{noisy_jerk_oscillator}. But the case of (thermal) white noise, e.g. given in inertial active dusty plasma particles \cite{Nosenko} will be included as a special case. 

\subsection{Experimental realizations}
\subsubsection{Macroscopic feedback experiment}
\begin{figure}[htbp]
    \includegraphics[width=\linewidth]{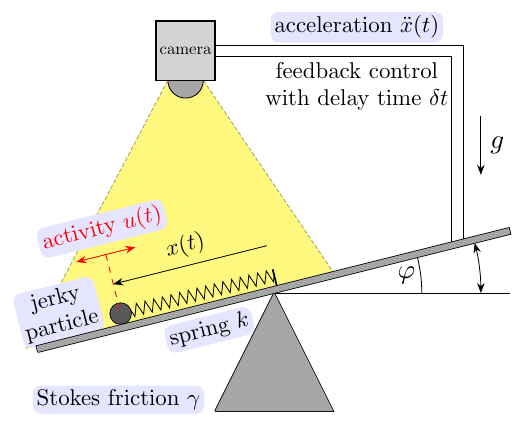}
    \caption{Schematic set-up to realize a jerky active particle in one dimension: The particle (black sphere) with bare mass $m_p$ is moving on a tilted plane and the tilt angle $\varphi$ is steered externally by delayed feedback resulting in a force $m_p g \sin (\varphi )$ that is an odd function of ${\ddot x}(t-\delta t)$ with $g$ denoting the gravitational acceleration and $\delta t > 0$ the time delay. In general, the particle is subject as well to a spring with spring constant $k$ and Stokes friction with a friction coefficient $\gamma$. Activity enters via active Ornstein-Uhlenbeck noise (shown in red) which makes the motion stochastic and can be programmed as well.}
    \label{fig:(1)}
\end{figure}

A macroscopic realization  of an active jerky particle is shown in Figure~\ref{fig:(1)}. This demonstrates that ``active jerks'' can conveniently be implemeted in experiments. Consider the one-dimensional motion of an inertial particle on a tilted plane where the tilt angle is controlled by external feedback: a camera measures the velocity change and translates this information into a change in the tilt angle. Typically this is done with a time delay $\delta t$ \cite{Cichos2}. The particle can be exposed to a spring force, to Stokes friction (e.g.\ by the surrounding air) and active noise. Hence, this set-up represents exactly the basic model \eqref{eq:(X0)}.

\subsubsection{Colloids with memory-dominated friction}
Here it is shown that the basic equation  \eqref{eq:(X0)} is also realized in  overdamped active colloids exposed to friction that includes memory, i.e.\ all the past particle velocities contribute to its total friction. Such effects do occur for active colloidal particles in non-Newtonian, viscoelastic solvents. For an active colloid in a harmonic trap of stiffness $k$, the equation of motion reads \cite{viscoelastic1,viscoelastic2,viscoelastic3}
\begin{equation}
    \label{eq:(colloids)}
    \int_{-\infty}^{\infty} \Gamma(t-t') \Theta(t-t'){\dot x}(t') \, dt' + k x(t) = f(t)
\end{equation}
where $\Gamma (t)\Theta(t)$ is the memory kernel in the friction and $f(t)$ is a generalized noise term, $\Theta(t)$ denotes the Heaviside step function. In Fourier space, the memory term factorizes and if we expand the Fourier transform of the memory kernel in Fourier space  as 
\begin{equation}
    \label{eq:(expand)}
    \int_0^{\infty} \exp( i \omega t) \Gamma(t) dt = \Gamma_0 +i\omega \Gamma_1 -\frac {\Gamma_2}{2} \omega^2 + O(\omega^3)
\end{equation}
with $\omega$ denoting the frequency and the moments $\Gamma_n = \int_0^{\infty} dt \, t^n \Gamma (t)$, we obtain the {\it same} equation of motion in Fourier space as in our basic model by identifying $\lambda$ with $\Gamma_2/2$, $m=\Gamma_1$ and $\gamma = \Gamma_0$. This demonstrates that our model is applicable also to active soft matter systems in viscoelastic environments and that the sign and magnitude of the effective mass and jerky coefficient can in principle be steered at wish for colloids in different responding backgrounds.

\subsection{Green's function and correlations}

Henceforth we assume that the particle is at complete rest at initial time $t=0$ such that $x(t=0)= {\dot x}(t=0) = {\ddot x}(t=0) =0$ and is then exposed to the stochastic and systematic forces for $t>0$. In the absence of activity ($v_0=0$), the general solution of the homogenous equation \cite{linear_jerk1,linear_jerk2} has been discussed before \cite{Linz_1998}. It is given by a superposition of exponentials $x(t)=\sum_{j=1}^3 A_j \exp(-i\omega_j t))$ where  $\omega_1$, $\omega_2$ and $\omega_3$  are the zeros of the cubic polynomial 
\begin{equation}
    \label{eq:(CC)}
    \omega^3 +i \frac{m}{\lambda} \omega^2 -  \frac{\gamma}{\lambda} \omega -i \frac {k}{\lambda} = (\ \omega-\omega_1)(\omega - \omega_2)(\omega-\omega_3).
\end{equation}
These three zeros can be determined by Cardano’s formula \cite{Cardan}, see Appendix~\ref{App:B} and further discussed in section IV. The amplitudes $A_j$ $(j=1,2,3)$ are complex and need to match the initial conditions at $t=0$. The dynamics of the solution of the homogeneous equation  depends crucially on the imaginary parts of the three zeros $\omega_1$, $\omega_2$, and $\omega_3$. Should at least one of these imaginary parts be positive, the general solution is exponentially growing in time and becomes unbounded.
In the following we consider the opposite case where all three imaginary parts of $\omega_1$, $\omega_2$, and $\omega_3$ are not positive, the requirements on the parameters $\lambda$, $m$, $\gamma$ and $k$ to fulfill that condition are given in \eqref{eq:(conditions2)} 
in Appendix~\ref{App:B}.  This leads to exponentially damped solutions for times $t>0$. For $k>0$, effects from the initial conditions at $t=0$ vanish for long times and the system runs into a steady state. The solution of the equation of motion \eqref{eq:(X0)} is then expressed by Fourier transform as
\begin{equation}
    \label{eq:(X14)}
    x(t) = \frac{1}{2\pi} \int_{-\infty}^{\infty} d \omega \frac{ \int_0^t dt' \gamma  u(t') \exp (-i\omega (t-t'))  }{ i\lambda \omega^3 - m \omega^2 - i \gamma \omega + k}.
\end{equation}

\begin{figure}[htbp]
    \centering
    \includegraphics[width=0.8\linewidth]{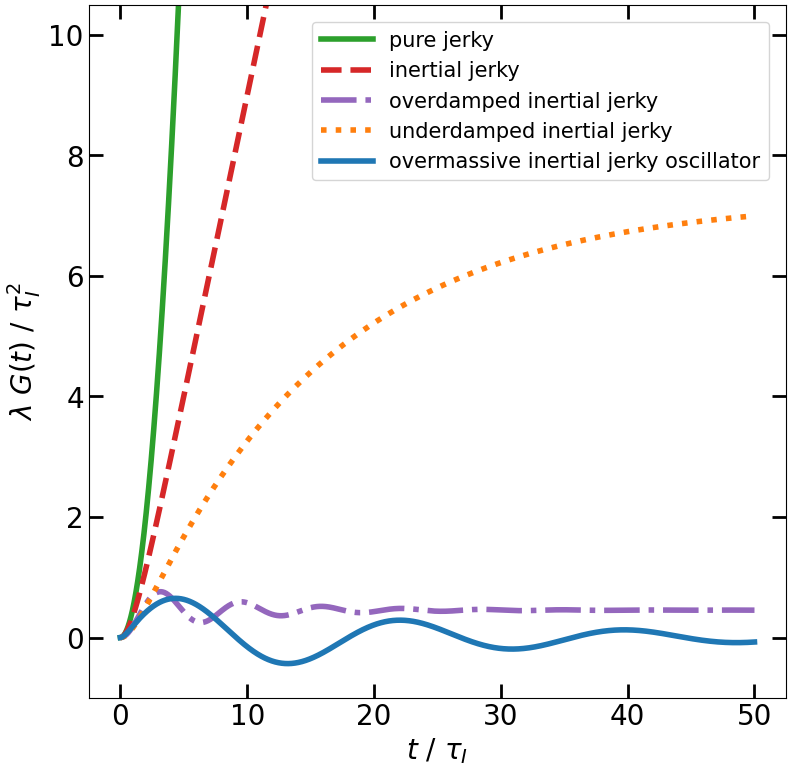}
    \caption{Green’s function of a jerky particle in units of $\tau_I^2/\lambda$ as a function of reduced time $t/\tau_I$ for the following cases: pure jerky (solid green) in arbitrary units, undamped inertial jerky (dashed red), overdamped inertial jerky (dot-dashed violet, for $\tau_f/\tau_I=1/4$) and underdamped inertial jerky (dotted orange, for $\tau_f/\tau_I=4$), jerky overmassive damped oscillator (solid blue for  $\tau_f/\tau_I=5.4$ and $\tau_s/\tau_I = 4.7$. 
    Only cases are shown where $G(t)$ does not increase exponentially in time. For small positive times $t$, $G(t)$ scales with $t^2$ and for $t<0$, $G(t)$ vanishes.
    }
    \label{fig:(2)}
\end{figure}

A characteristic quantity to discuss and classify active matter is the mean-square-displacement (MSD) of the particle \cite{Howse,Bechinger_RMP_2016} which governs its dynamical spreading. Since the initial position of the particle is vanishing, the MSD can be expressed as
\begin{equation}
    \label{eq:(X4)}
    \MSD (t) = \overline {(x(t)^2)} = 2\int_0^t dv \int_0^v dw M(v-w) G(v)G(w)
\end{equation}
 where $G(t)$ is  the Green´s function of \eqref{eq:(X0)} defined as 
\begin{equation}
    \label{eq:(general_green)}
    G(t) = \frac{1}{2\pi} \int_{-\infty}^{\infty} d \omega \frac{\exp (-i\omega t)}{ i\lambda \omega^3 - m \omega^2 - i \gamma \omega + k}.
\end{equation}
The latter can be analytically computed with the residue theorem such that the convolution with the noise kernel in Eqn.~\eqref{eq:(X4)} leads to the MSD. Clearly the Green’s function is completely decoupled from the noise, it is rather the response of the system to a $\delta$-kick $\delta(t)$ in time governed by the equation
\begin{equation}
    \label{eq:(G0)}
    \lambda {\dddot G}(t) + m {\ddot G}(t) + \gamma{\dot G}(t) + k {G}(t) = \delta (t).
\end{equation}
Examples for the Green’s function are summarized in Figure~\ref{fig:(2)} and will be discussed later in detail. Explicit results will be given for different special cases in the next section.

In the cases of large persistence, \eqref{eq:(X4)} simplifies to
\begin{equation}
    \label{eq:(X4p)}
    \MSD (t) = \gamma^2 v_0^2 \bigg(\int_0^t dv  G(v)\biggr)^2
\end{equation}
while for white ``passive'' noise
\begin{equation}
    \label{eq:(X4w)}
    \MSD (t) = 2\gamma^2 D \int_0^t dv \ (G(v))^2.
\end{equation}
Before proceeding further, it is useful to scale space and time with appropriate units. For $\lambda/m >0$, $ \lambda/\gamma>0$ and  $\lambda/k>0$,  the equation of motion \eqref{eq:(X0)} can be written as
\begin{equation}
    \label{eq:(X000)}
    {\dddot x}(t) + {\ddot x}(t)/\tau_I + {\dot x}(t)/{\tau_f}^2 + {x}(t)/{\tau_s}^3 = \gamma  u(t)/\lambda
\end{equation}
with four time scales: the persistence time $\tau_p$, the inertial time $\tau_I = \lambda/m$, the friction time $\tau_f = \sqrt{\lambda/\gamma}$ and the spring time $\tau_s = \sqrt[3]{\lambda/k}$. Henceforth we take if possible the inertial time $\tau_I$ as the natural time unit and $\gamma v_0\tau_I^3/\lambda$ as the natural length unit. Then a natural unit for the Green’s function is $\tau_I^2/\lambda$. Finally we remark that a {\it negative inertial mass\/} $m$ is formally possible within our analysis if all three conditions $\lambda<0$, $\gamma \leq 0$ and $k\leq 0$ are fulfilled.
One can also directly access the time-dependent mean kinetic energy $\Wkin(t)$ defined as
\begin{equation}
    \label{eq:(Wkin)}
    \Wkin (t) = {\frac {m}{2}}  \overline {({\dot x}(t)^2} = m\int_0^t dv \int_0^v dw M(v-w) {\dot G}(v){\dot G}(w).
\end{equation}
We finally introduce a further correlation function, the time-dependent {\it mean-square jerk} $\MSJ(t)$ as
\begin{equation}
    \label{eq:(jerk)}
    \MSJ(t) =   \overline {({\dddot x}(t))^2} = 2\int_0^t dv \int_0^v dw M(v-w) {\dddot G}(v){\dddot G}(w)
\end{equation}
where the passive, persistent and active cases can be distinguished by different kernels given by Eqns.~\eqref{eq:(M1)}, \eqref{eq:(M2)} and \eqref{eq:(M3)}.
\section{Special cases of active jerks}
In the sequel different special cases are discussed step-by-step. While the case $\lambda =0$ has been explored before \cite{colored,Risken,Haenggi}, we describe here the solution for the MSD for $\lambda \not= 0$. In principle, the sign of the jerky coefficient $\lambda$ can be both positive and negative. Table~\ref{tab:(I)} summarizes the results of the following subsections in terms of the short-time and long-time scaling exponents for the MSD. Table~\ref{tab:(II)} gives the crossover between different dynamical regimes for different separated time scales and Table~\ref{tab:(III)} provides  corresponding scaling exponents for the mean kinetic energy.
\begin{table}[htbp]
\centering
\begin{adjustbox}{max width=\linewidth}
\begin{tabular}{|c| c|c || c|c || c|c |}
    \hline
    \diagbox[width=15em,height=2.5em,dir=NW]{\hspace{0.8cm}\textbf{parameters}}{\raisebox{-0.7em}{\textbf{noise}}\hspace{0.5cm}} 
    & \multicolumn{2}{c}{\textbf{active}} 
    & \multicolumn{2}{c}{\textbf{persistent}} 
    & \multicolumn{2}{c|}{\textbf{passive}} \\
    \hline
    \textbf{time} & short & long & short & long & short & long \\
    \hline
    pure jerky  & 6 & 5 & 6 & 6 & 5 & 5 \\
    \hline
    inertial jerky ($\lambda/m>0$)  & 6 & 3 & 6 & 4 & 5 & 3 \\
    \hline
    inertial jerky ($\lambda/m<0$)   & -- & -- & -- & -- & -- & -- \\
    \hline
    damped inertial jerky   & 6 & 1 & 6 & 2 & 5 & 1 \\
    ($\lambda/m>0,\,\lambda/\gamma >0,\,k=0$)   &  &  &  &  &  &  \\
    \hline
    jerky noninertial damped oscillator   & -- & -- & -- & -- & -- & -- \\
    ($m=0$)   &  &  &  &  &  &  \\   
    \hline
    jerky inertial undamped oscillator   & -- & -- & -- & -- & -- & -- \\
    ($\gamma=0$)   &  &  &  &  &  &  \\     
    \hline
    jerky inertial damped oscillator   & 6 & 0 & 6 & 0 & 5 & 0 \\
    \hline
    damped inertial oscillator   & 4 & 0 & 4 & 0 & 3 & 0 \\
    ($\lambda = 0$)   &  &  &  &  &  &  \\
    \hline
    overdamped Brownian oscillator   & 2 & 0 & 2 & 0 & 1 & 0 \\
    ($\lambda = m = 0$)   &  &  &  &  &  &  \\
    \hline
\end{tabular}
\end{adjustbox}
\caption{Summary of the different dynamical exponents governing the MSD for short and long times for various special cases and active, persistent and passive noise. A $-$ symbol means that the MSD increases exponentially in time such that there is no such finite exponent. The harmonic oscillator with colored noise \cite{colored} and the overdamped harmonic oscillator (used in many applications see e.g.\ \cite{Dhontbook,Doi,twenty_years,Krueger}) are given as a reference for comparison.}
\label{tab:(I)}
\end{table}

\subsection{Pure jerky particle}
\begin{figure}[htbp]
    \centering
    \includegraphics[width=\linewidth]{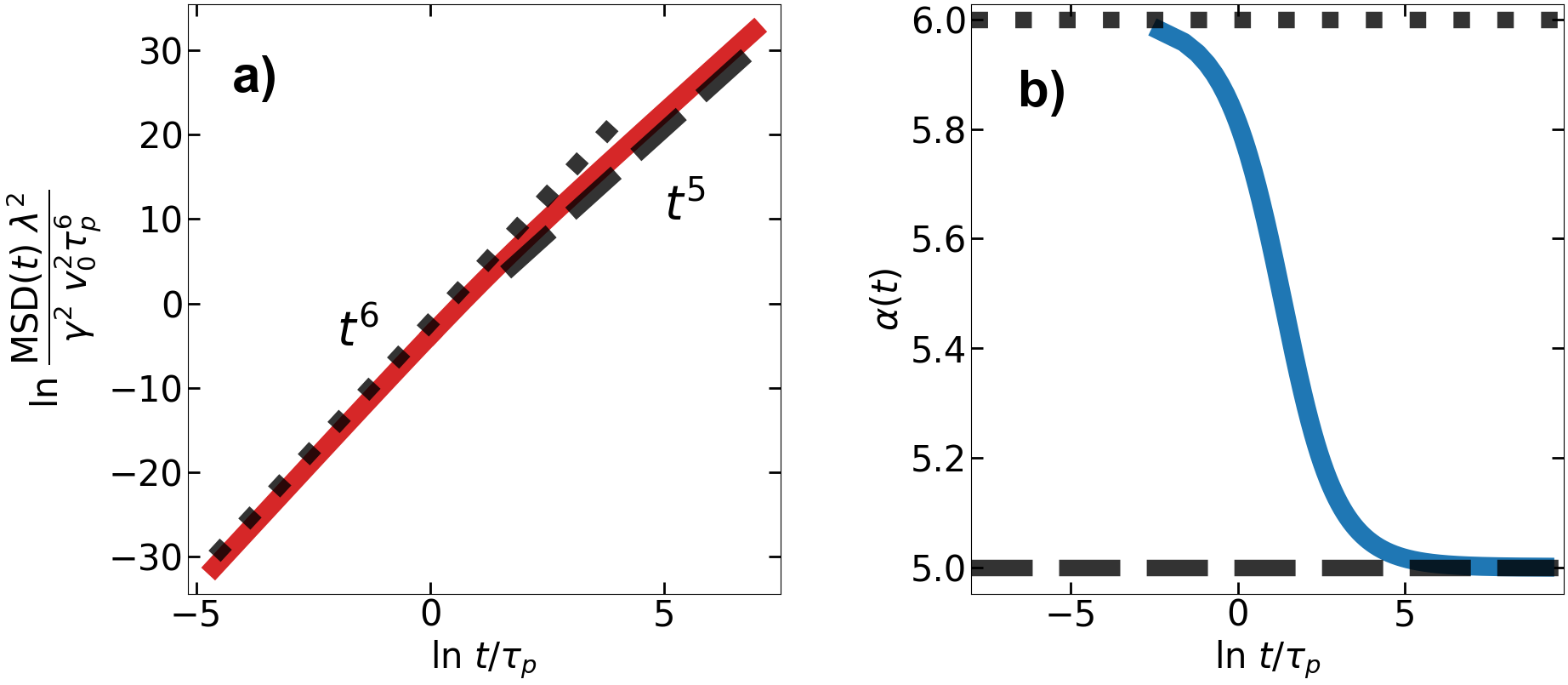}
    \caption{a) Mean-square displacement $\MSD(t)$  (in units of $\gamma^2 v_0^2\tau_p^6/\lambda^2$)  and b) dynamical exponent $\alpha (t)$ for a pure jerky particle as a function of $t/\tau_p$ for active (solid), persistent (dotted) and passive (dashed) noise.}
    \label{fig:(3)}
\end{figure}

Let us consider a pure jerky active particle, with zero mass $m=0$ and friction $\gamma=0$ freely moving in space such that $ k=0$.  A similar situation has been studied for telegraphic noise by Dean et al. \cite{Dean}. 
The Green’s function for a pure jerky particle is given by the scale-free expression
\begin{equation}
    \label{eq:(G00)}
    G(t) =\Theta(t)t^2/2\lambda
\end{equation}
shown in Figure~\ref{fig:(2)}. Hence the MSD for an active jerky particle is
\begin{equation}
\begin{aligned}
\label{eq:(X6)}
    \MSD(t) = {\frac {\gamma^2 v_0^2\tau_p^6}{\lambda^2}} \bigg[ &{\frac{(t/\tau_p)^5}{10}}  - {\frac{ (t/\tau_p)^4}{4}} + {\frac{(t/\tau_p)^3}{3}} -2 + \\
    + & \exp(-(t/\tau_p)) ((t/\tau_p)^2 + 2 (t/\tau_p)+ 2) \bigg]
\end{aligned}
\end{equation}
which is plotted on a double logarithmic scale in Figure~\ref{fig:(3)}. The MSD behaves asymptocially as   $ \gamma^2 v_0^2 t^6 / 36  \lambda^2$    for short times  and as $\gamma^2 v_0^2 \tau_p t^5 / 10 \lambda^2$  for large times, i.e.\ it involves a {\bf huge dynamical scaling exponent 6} for short times corresponding to a constant jerk (i.e.\ a constant time-derivative of an acceleration)  {\bf and 5} for long times. The crossover between the two dynamical regimes happens at a time of order $\tau_p$ which  is the only time scale involved here. The long-time exponent 5 is quite unusual in active matter \cite{shear,ten_Hagen_JPCM,Coulombfriction,Breoni,quantum} and describes an anomalous diffusion that is even faster than a constant acceleration. These huge exponents signify  an {\bf gigantic dynamical spread} very different from ordinary diffusion where this exponent is 1 or ballistic motion where the exponent is 2. 
In the limit of large persistence, the MSD reduces to  $ \gamma^2 v_0^2 t^6 / 36  \lambda^2\tau_p$    which is the short-time behavior introduced above while in the opposite limit of white noise,  $ \MSD (t) = \gamma^2 D t^5 / 10\lambda^2$ reflecting the long-time behavior from above. These results are included in Figure~\ref{fig:(3)}a for comparison.
In Figure~\ref{fig:(3)}b, we have included the dynamical exponent $\alpha (t)$ defined as a logarithmic derivative of the MSD 
\begin{equation}
    \label{eq:(E_alpha)}
    \alpha (t) = \totdiff{\ln (t)}{\ln \left(\MSD(t)\right)}. 
\end{equation}
The crossover in $\alpha(t)$ from 6 to 5 at the persistence time is clearly visible.


The dynamical exponents for the time-dependent mean kinetic energy $\Wkin(t)$ are summarized in Table~\ref{tab:(III)}. For the pure jerky particle they are  4 for short and 3 for long times.
To access the mean square jerk $\MSJ(t)$  via Eqn. \eqref{eq:(jerk)} we have ${\dddot G}(t)= \delta (t)/\lambda$. Therefore the mean square jerk is constant for $t>0$
\begin{equation}
    \label{eq:(constant)}
    \MSJ(t) = 2\gamma^2 v_0^2 \Theta (t) /\lambda^2
\end{equation}
for both active and persistent cases while this constant is diverging in the passive case. This is indicated in Figure~\ref{fig:(5)}. 
\subsection{Inertial jerky particle}

\begin{figure}[htbp]
    \centering
    \includegraphics[width=\linewidth]{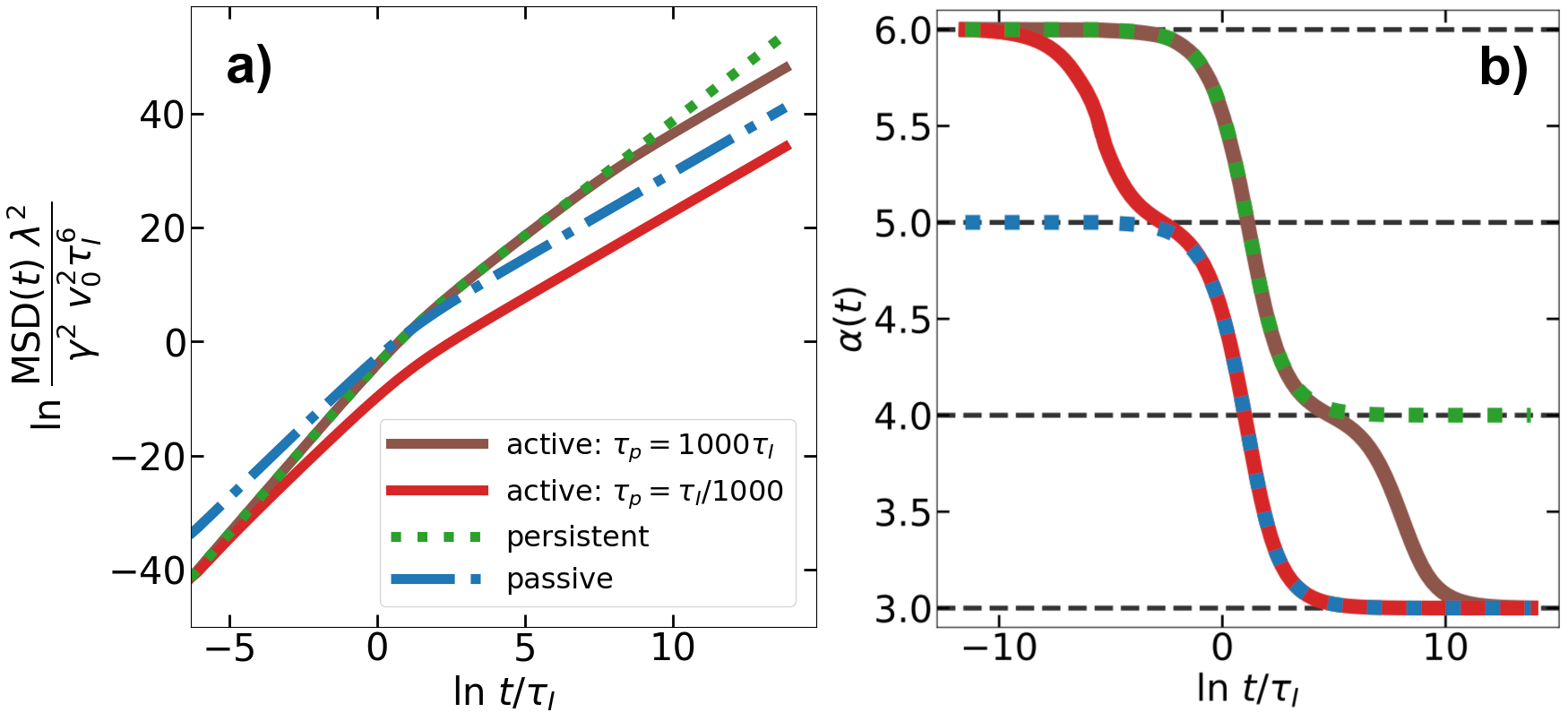}
    \caption{a) Mean-square displacement $\MSD(t)$ (in units of $\gamma^2 v_0^2\tau_I^6/\lambda^2$) and b) dynamical exponent $\alpha (t)$ for an inertial jerky particle as a function of $t/\tau_I$ for active (solid), persistent (dotted) and passive (dashed-dotted) noise: $\tau_p = \tau_I/1000$ (red) with crossovers $t^6\to t^5 \to t^3$ and $\tau_p = 1000 \tau_I$ (brown) with crossovers $t^6\to t^4 \to t^3$.}
    \label{fig:(4)}
\end{figure}

Next we compute the MSD for an undamped {\it inertial jerky particle} where $\gamma = k =0$. We consider the case $m/\lambda>0$, then the dynamics is such that an increase in acceleration is weakened by a jerk and vice versa stabilizing somehow the dynamics fulfilling the condition of bounded solutions in the homogeneous case, see also IV. Now two different time scales are entering, namely the persistence time $\tau_p$ and the jerky inertial time $\tau_I$. The Green’s function is  given by
\begin{equation}
    \label{eq:(G11)}
    G(t) = {\frac{\tau_I^2}{\lambda}}\Theta (t) [\,  t /\tau_I  - 1 + \exp (- t/\tau_I] )\,]    
\end{equation}
(see Figure~\ref{fig:(2)}) and the MSD  shown in Figure~\ref{fig:(4)}a is
\begin{equation}
\label{eq:(X7)}
\begin{aligned}
    &\MSD(t) = {\frac{2\gamma^2 v_0^2\tau_I^6}{\lambda^2}} \bigg[\frac{1}{3} \xi(t/\tau_I)^3 - \frac{1}{2} (2\xi + \xi^2) (t/\tau_I)^2 \\
    &+ (\xi + \xi^2) t/\tau_I + B I_1(t/\tau_I, 1/\xi) - B I_0(t/\tau_I, 1/\xi) \\
    &+ B I_0(t/\tau_I, 1+1/\xi) + ({\frac{\xi}{1-\xi}}+\xi) I_1(t/\tau_I, 1) \\
    &- (\xi + \xi^2 +{\frac{\xi}{1-\xi}}) I_0(t/\tau_I, 1)
    + {\frac{\xi}{1-\xi}}  I_0(t/\tau_I, 2) \bigg].
\end{aligned}
\end{equation}

Here we have introduced the integrals $I_0(x,\beta ) =\int_0^x \:  dy \: \exp (-\beta y) = [1-\exp (-\beta x )]/\beta$ and $I_1(x,\beta ) =\int_0^x \: dy \: y \: \exp (-\beta y) = [1-\exp (-\beta x )(\beta x +1)]/\beta^2$ as well as the constant $B=\xi^2 + \xi - \xi/(1-\xi)$ and the time ratio $\xi = \tau_p/\tau_I$. 
For short times, the MSD reduces again to $\gamma^2 v_0^2 t^6 /36\lambda^2$ and for large times to $2\gamma^2 v_0^2  \tau_p\tau_I^2 t^3/3 \lambda^2$ with exponents 6 and 3 respectively which are unusual, in particular in the long-time limit. For $\tau_I \simeq \tau_p$, a crossover from the short-time $t^6$ to the long-time $t^3$ scaling occurs roughly at this time. However, for $\tau_p\ll\tau_I$, there are two crossovers: the first crossover occurs at time $\tau_p$ from $t^6$ to $t^5$ scaling, then the second occurs at $\tau_I$ from $t^5$ to $t^3$ scaling. Conversely, for $\tau_p\gg\tau_I$, there are again two crossovers but now the first occurs at time $\tau_I$ from $t^6$ to $t^4$ scaling and the second happens at $\tau_p$ from $t^4$ to $t^3$ scaling revealing an exponent 4 for intermediate times. This is demonstrated in Figure~\ref{fig:(4)}b.

In the persistent case we get
\begin{equation}
    \label{eq:(YYY0)}
    \MSD(t) = {\frac {\gamma^2 v_0^2\tau_I^6} {\lambda^2 }} \bigg[\frac{1}{2} (t/\tau_I)^2 - t/\tau_I +1 - \exp (- t/\tau_I)\bigg]^2
\end{equation}
which again behaves as  $ \gamma^2 v_0^2 t^6 / 36  \lambda^2$    for short but as $ \gamma^2 v_0^2 \tau_I^2 t^4 / 4  \lambda^2$ for large times.  For white noise, on the other hand, the MSD for an inertial active jerky particle is given by
\begin{equation}
    \label{eq:(YYY)}
    \begin{aligned}
        \MSD(t) &= \frac {2\gamma^2 D\tau_I^5 }{\lambda^2} \bigg[  {\frac {1}{3}} (t/\tau_I)^3 - (t/\tau_I)^2 + t/\tau_I \\
        & + I_0(t/\tau_I,2) - 2 I_0( t/\tau_I,1) + 2I_1( t/\tau_I,1)\bigg].
    \end{aligned}
\end{equation}
Here, for short times, the MSD scales as $\gamma^2 v_0^2 \tau_p  t^5/5\lambda^2$ with the exponent 5 and for long times we have $4\gamma^2 D \tau_I^2 t^3/3\lambda^2$ with the exponent 3 such that the crossover happens at a time proportional to $\tau_I$.

We refer to Table~\ref{tab:(III)} for the dynamical exponents governing the time-dependent mean kinetic energy $\Wkin(t)$. With ${\dddot G}(t)= \delta (t)/\lambda - \Theta (t) \exp ( - t/\tau_I ) / \lambda \tau_I$ one obtains for the mean square jerk  
\begin{equation}
\label{eq:(Jerk_1)}
\begin{aligned}
\MSJ(t) &= {\frac{2\gamma^2 v_0^2}{\lambda^2}} \Theta (t) \bigg[ 1  -\frac{\xi}{2\xi -2}  (  1- \exp (-2t/\tau_I))
         \\ &+ {\frac {\xi}{ \xi^2 -1}}  (  1- \exp (-\frac{\xi+1}{\xi}    t/\tau_I)  )  \bigg] . 
\end{aligned}
\end{equation} 
 This is shown in Figure 5. For short times, $\MSJ(0) = {\frac{2\gamma^2 v_0^2}{\lambda^2}}$, which is the constant \eqref{eq:(constant)} obtained in the pure jerky case. For long times, the MSJ approaches another constant
\begin{equation}
    \label{eq:(xi)}
    \lim_{t\to \infty} \MSJ(t)  ={\frac{\gamma^2 v_0^2}{\lambda^2}} {\frac {\xi +2}{\xi +1}}.
\end{equation}
 For $\xi \to 0$, $\MSJ(t\to\infty)= 2\gamma^2 v_0^2/\lambda^2$. In the opposite limit of high persistency, $\xi\to\infty$, the MSJ approaches $\gamma^2 v_0^2/\lambda^2$. The  constant MSJ$(t\to\infty) $ is approached from below for large times except for $\tau_p/\tau_I = 0, 1, \infty$ and there is a minimum in MSJ$(t)$ which occurs at a time $t_0= \tau_p (\ln \xi )/(\xi - 1)$.
\subsection{Damped inertial jerky particle}
Third, we explore the case of a damped inertial jerky active particle where $k=0$ but $\lambda$, $m$ and $\gamma$ are kept as general as possible. For stability reasons, we restrict the treatment to $\tau_I=\lambda/m>0$ and $\lambda /\gamma>0$. There are two situations: i) the {\it   ``underdamped''  inertial jerky particle} where $\tau_f > 2\tau_I>0$ and ii) the {\it ``overdamped'' inertial jerky particle} where $0<\tau_f < 2\tau_I$. The general solution for $G(t)$ can be written as
\begin{equation}
    \label{eq:(GGG)}
    G(t) = {\frac{1}{\lambda}} \Theta (t)  {\frac {1}{(\omega_1-\omega_2)}}   \sum_{j=1}^2   (-1)^j \frac {1- \exp (-i\omega_j t)  }{\omega_j}
\end{equation}
where the two eigenfrequencies are
\begin{equation}
    \label{eq:(eigen)}
    \omega_{1,2} = -{\frac {i}{2\tau_I}} \pm \sqrt{{ - \frac{1}{4 \tau_I^2}} + {\frac {1}{\tau_f^2}}} \:.
\end{equation}





Again the short-time part is unaffected by the damping, but for long times $G(t)$ now tends to a constant. Note that the role of damping for a jerky inertial particle is reversed relative to that of the traditional inertial harmonic oscillator: while small damping (underdamping) leads to oscillatory behavior of $G(t)$ in the latter case, large damping (overdamping) is needed to obtain oscillatory behavior in the former case.

The MSD is computed as 
\begin{equation}
    \begin{aligned}
        \MSD(t) =& \frac { 2\gamma^2 v_0^2}{\lambda^2} \biggl(  \frac {E(t,0,0)}{\omega_1^2\omega_2^2} + \sum_{j,\ell=1}^2   \frac {{(-1)^{j+\ell}} 
         E(t,i\omega_j,i\omega_\ell) } {(\omega_1-\omega_2)^2\omega_j\omega_\ell} \\+&\sum_{j=1}^2 \frac {(-1)^j ( E(t,i\omega_j,0)+E(t,0,i\omega_j))} {(\omega_2-\omega_1)\omega_1\omega_2\omega_j} 
      \biggr).
    \end{aligned}
\end{equation}
Here we have introduced the function
\begin{equation}
    \label{eq:(EE)}
    \begin{aligned}
        E&(t,\beta,\gamma) = \int_0^t dv \int_0^v dw \ \exp( - v/\tau_p + w/\tau_p -\gamma w - \beta v) \\
        =&\frac{1}{\beta - \omega_p} \bigg(
        \frac{1-\exp (-( \gamma + \omega_p)  t )   }{  \gamma + \omega_p  } -   \frac{1-\exp (-(\gamma + \beta )t) }{  \gamma + \beta }    \bigg) 
    \end{aligned}
\end{equation}
with the persistence rate $\omega_p = 1/\tau_p$ and the arguments $\beta$ and $\gamma$ being arbitrary complex numbers. The short-time scaling is unaffected but for long times the asymptotic scaling is now diffusive like $\MSD(t) \simeq 2\gamma^2 v_0^2 G(\infty)^2 t/\tau_p$ with $G(\infty)= -1/\lambda \omega_1\omega_2$. 
For persistent noise, the long-time dynamics gives a ballistic scaling with $\MSD(t) \simeq \gamma^2 v_0^2t^2G(\infty)^2/4$ while for white noise  the long-time scaling is diffusive with $\MSD(t) \simeq \gamma^2 D G(\infty)^2 t$, see Table~\ref{tab:(I)}.

Let us then discuss possible crossovers in the dynamical scaling. Here, for $\lambda >0$, we have three independent time scales: the persistence time $\tau_p$, the jerky inertial time $\tau_I$, and the frictional time $\tau_f$. There are 6 different cases for two complete separations between two out of these three time scales. The corresponding cases and computed crossover scalings are summarized in Table~\ref{tab:(II)}. They all start with $t^6$ for very small times and end with $t$ for very long times but for intermediate time windows various other scalings emerge.

\begin{figure}[htbp]
    \centering
    \includegraphics[width=0.8\linewidth]{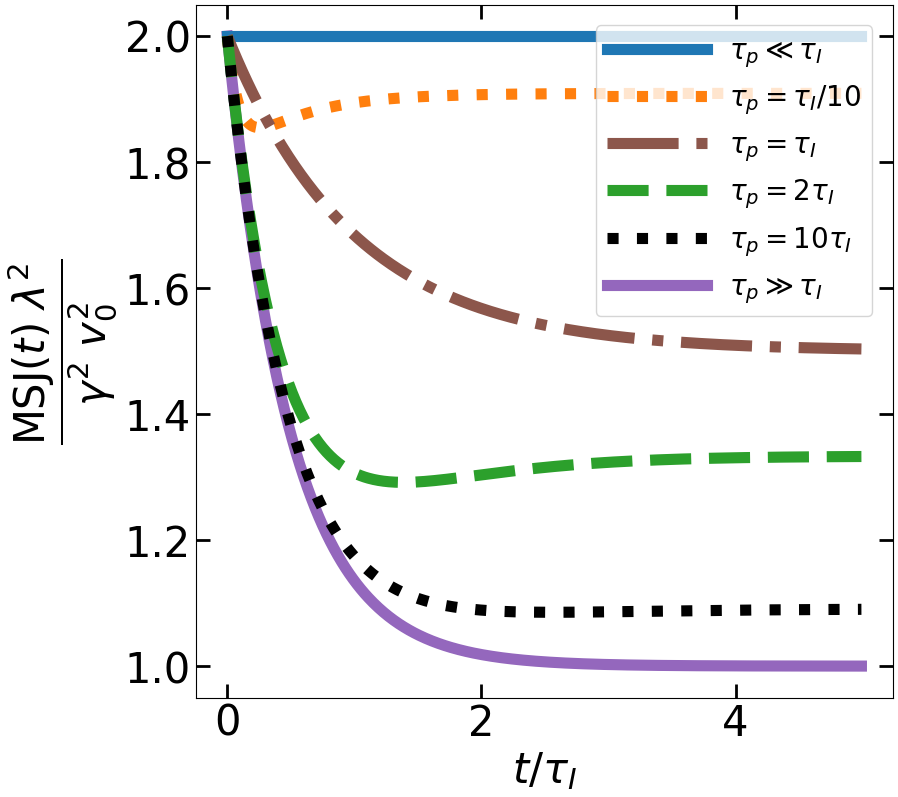}
    \caption{a) Mean-square jerk $\MSJ(t)$  (in units of $\gamma^2 v_0^2/\lambda^2$)  as a function of $t/\tau_I$ for an inertial  active particle for  $\tau_p = \tau_I/10$ (dotted orange), $\tau_p=\tau_I$ (dash-dotted brown), $\tau_p = 2\tau_I$ (dashed green), $\tau_p = 10 \tau_I$ (dotted black) and the two limits $\tau_p \ll \tau_I$ (blue) and $\tau_p \gg \tau_I$ (violet).}
    \label{fig:(5)}
\end{figure}

The mean-squared  jerk is
\begin{equation}
    \label{eq:(Jerk_2)}
    \MSJ(t) =  \frac{2\gamma^2 v_0^2}{\lambda^2} \Theta (t)        \sum_{j,\ell=1}^2  \frac {i(-1)^{j+\ell} \omega_ j^2 \omega_\ell^2  E(t,i\omega_j,i\omega_\ell)}{(\omega_1 - \omega_2)^2} 
\end{equation}
and oscillates in the overdamped case exponentially approaching a constant for large times. In the underdamped case the approach is purely exponential in time as for the inertial jerky case shown in  Figure~\ref{fig:(5)}.

A final remark is in order: the distributions for $\dot x$, $\ddot x$ and $\dddot x$ in the steady state are Gaussian and can directly be obtained from that of the colored harmonic oscillator \cite{colored,Bonilla} by replacing (resp.\ transforming) positions and velocities of the colored harmonic oscillator with $\dddot x$ and $\ddot x$ accordingly. The same transformation applies to any other dynamical correlations of $\dot x$, $\ddot x$ and $\dddot x$. Therefore the function $\MSJ(t)$ is basically the time-dependent mean squared acceleration of a colored harmonic oscillator with vanishing initial position and velocities. The latter function has only rarely been considered and studied. 
\begin{table}[htbp]
\centering
\begin{adjustbox}{max width=\linewidth}
\begin{tabular}{| c|c |}
    \hline
    \textbf{time scale separation} & 
    \begin{tabular}[c]{@{}c@{}} \textbf{sequence of dynamical} \\ \textbf{scaling exponents} \end{tabular} \\
    \hline
    $\tau_I \ll \tau_f \ll \tau_p$ & $6\to 4\to 2\to 1$ \\
    \hline
    $\tau_f \ll \tau_I \ll \tau_p$ & $6\to 2\to 2\to 1$ \\
    \hline
    $\tau_p \ll \tau_f \ll \tau_I$ & $6\to 5\to 1\to 1$ \\
    \hline
    $\tau_p \ll \tau_I \ll \tau_f$ & $6\to 5\to 3\to 1$ \\
    \hline
    $\tau_f \ll \tau_p \ll \tau_I$ & $6\to 2\to 1\to 1$ \\
    \hline
    $\tau_I \ll \tau_p \ll \tau_f$ & $6\to 4\to 3\to 1$ \\
    \hline
\end{tabular}
\end{adjustbox}
\caption{Possible crossovers for the active damped inertial jerky particle. For short times the scaling is always with $t^6$ and for long times it is diffusive with $t$. 6 different complete time scale separations of $\tau_p$, $\tau_I$, $\tau_f$ are shown and the four different scaling exponents $\alpha$ are listed. The interpretation of Table is as follows. For example, for the first line where $\tau_I\ll\tau_f\ll\tau_p$ there are four scaling exponents $6\to 4 \to 2 \to 1$ with three crossover times. In detail, there is a first crossover from $t^6$ to $t^4$ scaling at $t=\tau_I$, a second crossover from $t^4$ to $t^2$ scaling at $t=\tau_f$, and a third crossover from $t^2$ to $t$ scaling at $t=\tau_p$. If subsequent exponents are identical, only the prefactors of the scaling laws differ.}
\label{tab:(II)}
\end{table}

\subsection{Jerky noninertial damped harmonic oscillator}
In order to get insight into the action of jerky dynamics it is instructive to consider the jerky harmonic oscillator under damping ($k>0, \gamma > 0)$ but in the complete absence of inertia such that $m=0$. For this {\it noninertial damped jerky harmonic oscillator} it can be easily shown (see Appendix~\ref{App:B}) that for {\bf any} $\lambda\not= 0$ there is {\bf always} a zero of the cubic equation \eqref{eq:(CC)} with a positive imaginary part, which gives an exponential growing solution. Therefore the MSD and MSJ are diverging for long times, such that there are no finite dynamical exponents, see Table~\ref{tab:(I)}. It is remarkable that a particle with jerky dynamics can always escape from a harmonic potential, i.e. a harmonic potential is never strong enough to keep the particle localized. This is in marked contrast to $\lambda =0$ with $m=0$ representing the classical textbook example of an overdamped oscillator which leads to a finite MSD and MSJ for long times such that the particle is always localized. This counterintuitive result illustrates the desastrous action of jerk: jerky dynamics tends to spread particles enormously. As we shall show in section IV, a finite mass $m$, however, helps to keep the particle localized in a harmonic potential provided the jerk coefficient $\lambda$ is small enough.

\subsection{Jerky inertial undamped harmonic oscillator}
There is another limit where jerk leads to delocalize a particle from a harmonic trap, this is the jerky inertial undamped harmonic oscillator characterized now by $\gamma =0$. For vanishing jerk $\lambda =0$ this is the traditional undamped harmonic oscillator which possesses pure oscillatory solutions for vanishing noise. As shown in Appendix~\ref{App:B}, any small but finite jerk induces a solution which grows exponentially in time leading to a {\it complete delocalization} of the particle, i.e. the particle leaves the confining harmonic cage just by jerk. Again this demonstrates that the role of jerk enormously enhances the spreading dynamics.

\section{Active jerky inertial damped harmonic oscillator}
\subsection{Jerky inertial damped harmonic oscillator}
Before we consider the case of active noise, let us study the noise-free limit ($v_0 =0$) assuming throughout this section positive $\gamma$ and positive $k$. Then according to the Ostrogradski theorem \cite{Ostrogradski}, a strong reduction of stable solutions is expected \cite{Nesterenko}. In detail,
 the three different complex eigenfrequencies $\omega_1$, $\omega_2$, $\omega_3$ of the homogeneous equation of motion can be analytically obtained as zeros of the characteristic cubic polynomial \eqref{eq:(CC)} by Cardano's formula, see Appendix~\ref{App:B}. Either all three $\omega_j$'s have no real part (i.e.\ all eigensolutions are exponentially decaying in time), then we call the system {\it undermassive}. Or two frequencies have the same real part, then their eigensolutions have an oscillatory component and we call it {\it overmassive}. In this overmassive case, the imaginary part of the oscillating solutions is {\it always} larger than that of the pure exponential solution such that for long times the oscillation is dominating. The transition line separating the undermassive and overmassive regions can be computed analytically. By rescaling position and time appropriately one finds that there are only two independent dimensionless parameters which characterize the qualitative behavior of the $\omega_j$'s which are the {\it dimensionless jerk} $\tilde{\lambda} = \lambda \gamma^3/k^2 $  and the {\it dimensionless mass} $\tilde{m} =\gamma^2/k$ such that the full behavior can be shown in the $\tilde{m}\tilde{\lambda}$-plane, see Figure~\ref{fig:(6)}. There are two branches of lines $\tilde{\lambda}_\pm(m)$ dividing the undermassive and the overmassive region which are given by
\begin{equation}
    \label{eq:massive}
\tilde{\lambda}_\pm({\tilde{m}})= \frac {\tilde{m}}{3} - \frac {2}{27} \pm \frac{2}{\sqrt{27} } ( \frac {1}{3} - {\tilde{m}})^{3/2} .  
\end{equation}
Another characteristic mass is that associated with the transition from overdamped to underdamped case of the traditional harmonic oscillator (at $\lambda=0$) which happens at $\tilde{m}=1/4$.
For ${\tilde{m}}< 1/4$, $\tilde{\lambda}_+(m) $ is positive, except at $\tilde{m}=0$ where $\tilde{\lambda}_+(m) =  \tilde{m}^2/2 + O(\tilde{m}^3)$ is vanishing quadratically in $\tilde{m}$.  For $1/4<\tilde{m}<1/3$  there are two positive solutions
 $\tilde{\lambda}_-= \tilde{m}-1/4 + O((\tilde{m}-1/4)^2)$ and $\tilde{\lambda}_+= 1/54 - 46 (\tilde{m}-1/4)/3 +  O((\tilde{m}-1/4)^2) $ and beyond the threshold $\tilde{m}=1/3$  real solutions for ${\tilde{\lambda}}_\pm(\tilde{m})$ do not exist. Hence for large ${\tilde{m}}$ we are always in the oscillating overmassive regime. At $\tilde{m}=1/3$ and $\tilde{\lambda}= 1/27$, there is a {\it cusp-like singularity}, see again Figure~\ref{fig:(6)}. The undermassive region continues and opens up for negative masses $\tilde{m}<0$, scaling asymptotically as $\tilde{\lambda}_\pm = \pm 2| \tilde{m} |^{3/2}/\sqrt{27}$ for $ \tilde{m}\to -\infty$. It is worth to note that there is a {\it double reentrance effect} as visualized in the cusp-like singularity of the undermassive region: for fixed small and positive $\tilde{\lambda}$, as $\tilde m$ is increased, one obtains the sequence undermassive $\to$ overmassive $\to$ undermassive $\to$ overmassive.

If all three eigenfrequencies $\omega_1$, $\omega_2$, $\omega_3$ are distinct and possess a non-positive imaginary part,  the Green's function is given by
\begin{equation}
    \label{eq:(BIG)}
    \begin{aligned}
            G(t) =& -\frac {1}{\lambda} \Bigg[       \frac{\exp (-i\omega_1 t)}{(\omega_1 - \omega_2)(\omega_1 - \omega_3)}  +  \frac{\exp (-i\omega_2 t)}{(\omega_2 - \omega_1)(\omega_2 - \omega_3)}  \\
            &+  \frac{\exp (-i\omega_3 t)}{(\omega_3 - \omega_1)(\omega_3 - \omega_2)}                    \Bigg] \\
            =& -\frac {1}{\lambda} \sum_{j,\kappa,\ell=1; \kappa<l}^3     |\epsilon_{j\kappa\ell}| \frac{\exp (-i\omega_j t)}{   (\omega_j -\omega_\kappa)(\omega_j - \omega_\ell )} 
    \end{aligned}
\end{equation}
where $\epsilon_{j\kappa\ell}$ denotes the Levi-Civita tensor.
Again $G(t)$ scales quadratically in time for small times and tends to zero for large times. In the overmassive region it always decays in an oscillatory fashion as a function of time, since the decay time of the oscillatory solutions is always smaller than that of the pure exponential solution what can be deduced from an analysis of the three $\omega_j$'s.

If at least one imaginary part of the three eigenfrequencies is positive, the Green's function will grow exponentially in time proportional to $\exp(\mu t)$ with $\mu=$ Max $( Im (\omega_1), Im (\omega_2), Im (\omega_3))$ where $Im$ denotes the imaginary part of a complex number. The conditions to avoid any exponential growth in time are given in Appendix~\ref{App:B}. If the conditions are not fulfilled, the quantities $\MSD (t) $, $\Wkin(t)$ and $\MSJ (t)$  will diverge exponentially in time proportional to $\exp(2\mu t)$. Depending on the real parts of the three eigenfrequencies $\omega_j$, this is either an envelope with an oscillatory behavior called {\it giant breathing} \cite{twenty_years} or a pure exponential growth, i.e.\ an {\it explosion}. Let us remark that giant breathing does not occur for the traditional damped oscillator ($\lambda = 0$) which shows even for negative mass $m$ and positive $\gamma$ and $k$ only explosions for unstable solutions. We shall discuss the consequences for an underlying localization-delocalization transition in the sequel.

\subsection{Relaxation towards the steady state}

Let us first discuss the time-dependent relaxation of the $\MSD (t) $, $\Wkin(t)$ and $\MSJ (t)$ towards the steady state.  In the following we use the compact  notation
\begin{equation}
    \mathbf{\sum}^{'} (\ldots) = \frac {2\gamma^2 v_0^2}{\lambda^2} \sum_{j,\kappa,\ell=1; \kappa<l}^3 \  \sum_{\nu,\beta,\sigma=1; \beta<\sigma}^3  | \epsilon_{j\kappa\ell}  \epsilon_{\nu \beta\sigma} | (\ldots) \ .
\end{equation}
 This double sum involves nine non-vanishing terms. Then the $\MSD (t)$ can be expressed using \eqref{eq:(X4)} and \eqref{eq:(general_green)} as
\begin{equation}
  \label{eq:(MSD_BIG)}
  \MSD(t) = \mathbf{\sum}^{'} \frac {   E(t,i\omega_\nu, i\omega_j  )  } {(\omega_j - \omega_\kappa)(\omega_j - \omega_\ell )  (\omega_\nu - \omega_\beta)(\omega_\nu- \omega_\sigma )}.
\end{equation}
The short-time scaling is unaffected proportional to $t^6$ 
Now four crossovers between different regimes of the MSD are possible in 24 different situations. For example, for $\tau_p\ll\tau_I \ll\tau_f\ll\tau_s$ the sequence of the scaling exponents in the spirit of Table~\ref{tab:(II)} is $6\to 5\to 3\to 1 \to 0$ etc.

For the sake of completeness we also give the results for the MSD in the persistent and passive limits. 
For persistent noise 
\begin{equation}
    \label{eq:(general_persistent)}
    \MSD(t)   = \mathbf{\sum}^{'} \frac { (\exp (-i\omega_j t) -1) (1-\exp ( -i\omega_\nu t) )}  {\omega_j\omega_\nu(\omega_j - \omega_\kappa)(\omega_j - \omega_\ell )  (\omega_\nu - \omega_\beta)(\omega_\nu- \omega_\sigma )}
\end{equation}
while in the passive case we obtain

{\footnotesize
\begin{equation}
\label{eq:(general_passive)}
    \MSD(t) = \frac{iD}{v_0^2} \mathbf{\sum}^{'} \frac { \exp (-i(\omega_j +\omega_\nu)t) -1  }  {(\omega_j - \omega_\nu)(\omega_j - \omega_\kappa)(\omega_j - \omega_\ell )  (\omega_\nu - \omega_\beta)(\omega_\nu- \omega_\sigma )}.
\end{equation}
}

The dynamical scaling exponents of the mean kinetic energy are given in Table~\ref{tab:(III)}. The analytical expression for $\Wkin(t)$ is given by 
\begin{equation}
    \label{eq:(general_Wkin)}
    \Wkin(t)   = \frac{m}{2}\mathbf{\sum}^{'} \frac { -\omega_j\omega_\nu  E(t,i\omega_\nu, i\omega_j  )  } {(\omega_j - \omega_\kappa)(\omega_j - \omega_\ell )  (\omega_\nu - \omega_\beta)(\omega_\nu- \omega_\sigma )}.
\end{equation}
For the mean squared jerk we find
\begin{equation}
    \label{eq:(general_jerk)}
    \MSJ(t) = \mathbf{\sum}^{'}  \frac {   -\omega_j^3 \omega_\nu^3 E(t,i\omega_\nu, i\omega_j    )} {(\omega_j - \omega_k)(\omega_j - \omega_\ell )  (\omega_\nu - \omega_n)(\omega_\nu- \omega_\sigma )}.
\end{equation}
Depending on the three eigenfrequencies $\omega_j$ and on $\tau_p$, the approach of the MSD$(t)$, $\Wkin(t)$ and MSJ$(t)$ to their steady state limits is either purely exponential or oscillatory with an exponential envelope in time $t$. Oscillating MSD's are sometimes called {\it breathing} particles \cite{twenty_years}. The breathing property is independent of the self-propulsion velocity $v_0$ and the persistence time $\tau_p$ but the breathing frequency itself depends on $\tau_p$ as can be deduced from \eqref{eq:(EE)}.

\subsection{Localization-delocalization transition}
For long times, correlations reflect those in the steady state.  In this limit, the MSD saturates at a constant $a^2  = \MSD(t\to\infty)$ , indicating a finite smearing of the particle in the harmonic potential with a spatial extent $a$. For an active particle without jerk, the spread $a$ is an important parameter to characterize the strength of the activity, see e.g.\ \cite{Szamel,Buttinoni_Caprini_Lowen_EPL,Szamel2}. 
Formally the dynamical exponent $\alpha$ vanishes for long times. The spatial extent $a>0$ can be computed as

\begin{figure*}[htbp]
    \centering
    \includegraphics[width=\linewidth]{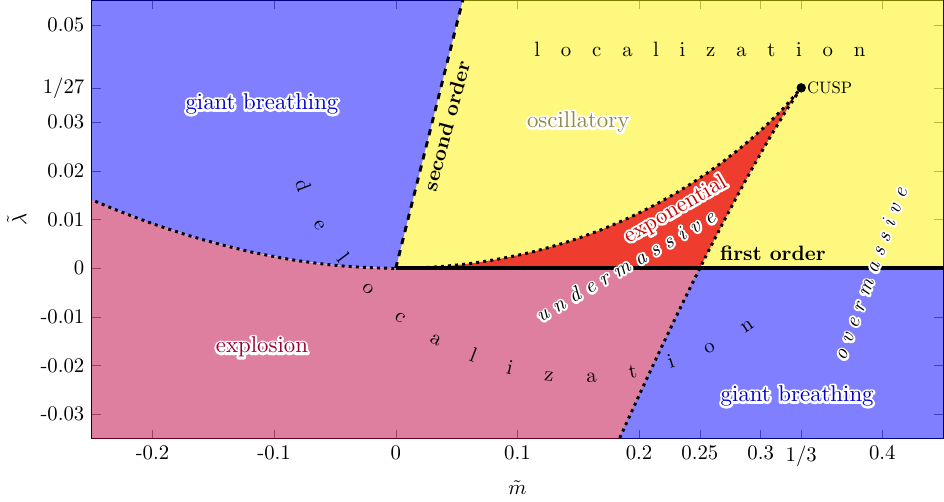}
    \caption{State diagram and relaxational dynamics of a jerky harmonic oscillator  as a function of the dimensionless jerk  $\tilde \lambda$ and the dimensionless mass $\tilde m$. As a function of $\tilde \lambda$ there are both first-order and second-order localization-delocalization transitions in the steady state. The first order transition line $\tilde{\lambda}(\tilde{m})$ is shown as a solid curve (at $\tilde{\lambda}(\tilde{m}) =0$ for $\tilde{m}>0$), the second order transition line $\tilde{\lambda}_c(\tilde{m})$ as a dashed curve. The separation lines $\tilde{\lambda}_\pm (\tilde{m})$ between the undermassive and overmassive regions are shown as dotted curves. Together with the localization region they define four different relaxational processes: exponential approach to the steady state (red region), exponentially damped but oscillatory approach to the steady state (yellow region), giant breathing which is exponentially growing and oscillatory in time (blue region) and an explosion which is a pure exponential divergence (violet region). 
    }
    \label{fig:(6)}
\end{figure*}

{\footnotesize
\begin{equation}
    \label{eq:(general_a)}
    a^2  = \mathbf{\sum}^{'}      \frac {   1  }      {(\omega_j - \omega_\kappa)(\omega_j - \omega_\ell )  (\omega_\nu - \omega_\beta)(\omega_\nu- \omega_\sigma ) (i\omega_p - \omega_j) (\omega_j + \omega_\nu)}
\end{equation}
}
and its inverse is plotted in Figure~\ref{fig:(7)}a. Here it can be concluded that jerk favors spreading and that the impact of jerk is enormous. There can be even a {\bf jerk-induced transition from localization} ($a<\infty$) {\bf to delocalization} where $a$ diverges exponentially in time to infinity. The transition point does not depend on the activity but solely on the two dimensionless parameters $\tilde \lambda$ and $\tilde m$. Figure~\ref{fig:(6)} shows the localization-delocalization transition in the $\tilde m \tilde\lambda$-plane. The transition line is determined by the conditions \eqref{eq:(conditions2)}. 
A characteristic order parameter $\phi$ for the localization-delocalization transition is the inverse localization length $\phi = 1/a^2$. In the spirit of this order parameter the localization-delocalization transition is first order if the order parameter jumps as the parameters are varied or second order if it is continuous but if the first derivative of the order parameter with respect to the system parameters is discontinuous.
In fact both transition types do occur, a first order transition line as well as a second order transition line. For $\tilde m>0$, the first order transition is given by
\begin{equation}
    \label{eq:crit1}
\tilde \lambda (\tilde m) =0 
\end{equation}
while a line of second order transitions occurs at $ \tilde \lambda_c(\tilde m)$.  For small positive mass $\tilde m$, we obtain the linear relation 
\begin{equation}
    \label{eq:crit2}
\tilde \lambda_c(\tilde m) = \tilde m + O(\tilde{m}^2).
\end{equation}
For large $\tilde m$, on the other hand,  the critical line is given by 
\begin{equation}
    \label{eq:crit3}
\tilde \lambda_c(\tilde m) = \frac {2\tilde m^{3/2}}{\sqrt{37}} + \frac{3\tilde m}{74 }  + O(\sqrt{\tilde m})
\end{equation}
which scales with dimensionles mass $\tilde m^{3/2}$ involving a nontrivial scaling exponent $3/2$. The stability region is plotted in Figure~\ref{fig:(6)}. For negative mass $\tilde m$ there is always delocalization. Interestingly, while typically increasing the spring constant $k$ (for fixed $\gamma$, $m$ and $\lambda$) leads to localization, under the action of jerk the effect is reversed: increasing the spring constant $k$ favors delocalization.

For long times, in the steady state, the mean kinetic energy $\Wkin(t\to\infty)$ saturates at the following constant
{\footnotesize
\begin{equation}
\label{eq:(general_Wkin_saturation)}
\frac{m}{2}\mathbf{\sum}^{'}  \frac {   -\omega_j\omega_\nu }      {(\omega_j - \omega_\kappa)(\omega_j - \omega_\ell )  (\omega_\nu - \omega_\beta)(\omega_\nu- \omega_\sigma ) (i\omega_p - \omega_j) (\omega_j + \omega_\nu)   } 
\end{equation}
}
and the averaged jerk turns to the following finite constant as $t\to \infty$: 
{\footnotesize
\begin{equation}
\label{eq:(general_jerky)}
\mathbf{\sum}^{'}  \frac {   -\omega_j^3\omega_\nu^3 }      {(\omega_j - \omega_\kappa)(\omega_j - \omega_\ell )  (\omega_\nu - \omega_\beta)(\omega_\nu- \omega_\sigma ) (i\omega_p - \omega_j) (\omega_j + \omega_\nu)} . 
\end{equation}}
\noindent

\subsection{Effective temperatures}

\begin{figure}[htbp]
    \centering
    \includegraphics[width=\linewidth]{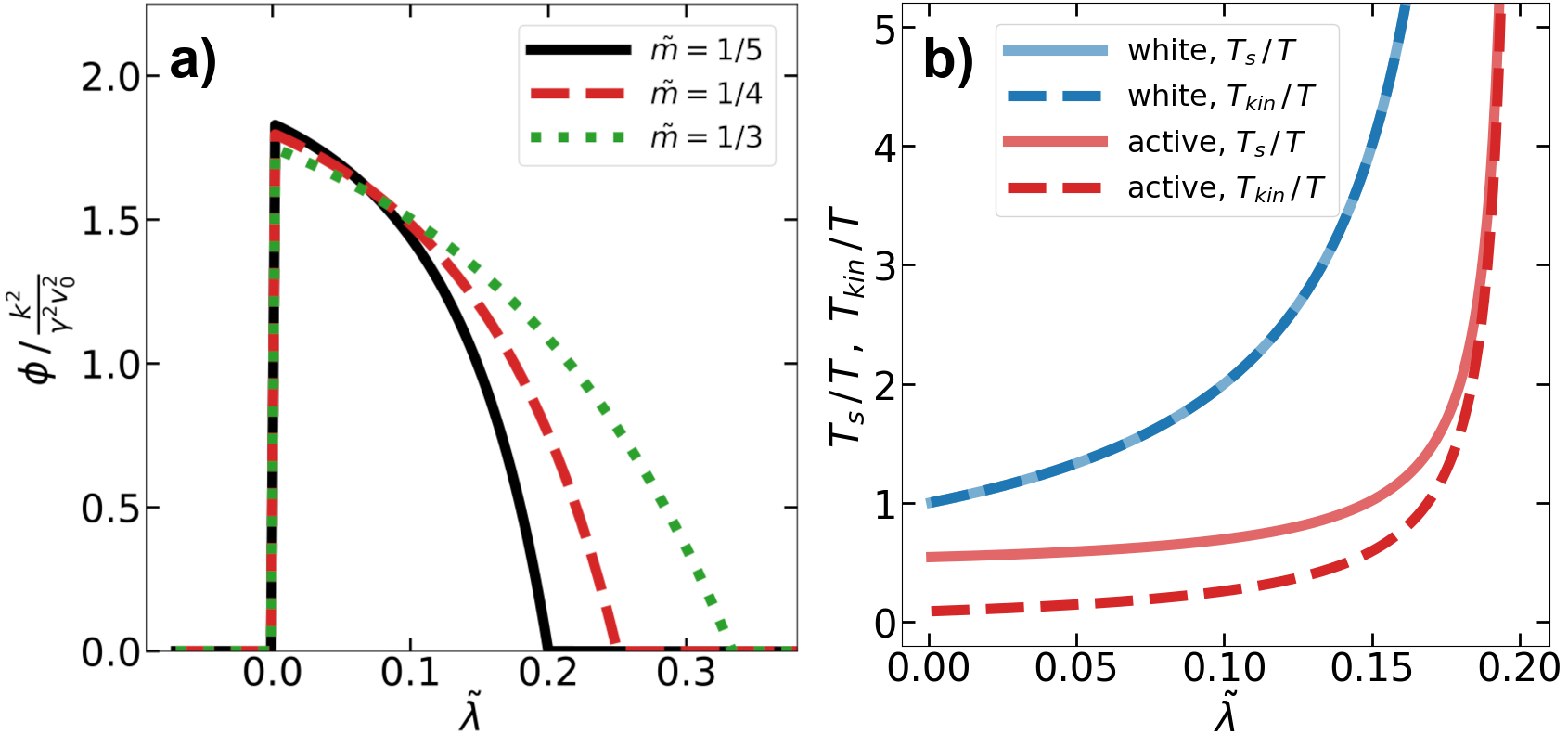}
    \caption{a) Order parameter $\phi$ of the localization-delocalization transition: Inverse mean-squared position $1/a^2$ (in units of $k^2/\gamma^2 v_0^2$) in the steady state as a function of   $\tilde\lambda$ for $\tau_p = \tau_I$ and three different values for $\tilde m$: $\tilde m = 1/5$ (solid line), $\tilde m = 1/4$ (dashed line), $\tilde m = 1/3$ (dotted line). b) Effective temperature ratios $T_s/T$ (solid line) and $\Tkin/T$ (dashed line) for $\tilde m = 1/5$ as a function of the dimensionless jerk parameter $\tilde \lambda$ for white noise (red) and active noise (blue) with $\tau_p=\tau_I$.}
    \label{fig:(7)}
\end{figure}

In nonequilibrium there are various ways to define a temperature which all coincide in equilibrium \cite{Szamel,Hecht_JCP}.   For white noise, the fluctuation-dissipation theorem gives $T=\gamma D / k_B$ where $k_B$ is Boltzmann's constant. For active noise this can be generalized to
\begin{equation}
\label{eq:(standard_temperature)}
T = \gamma v_0^2 \tau_p / k_B .
\end{equation}
\noindent
Moreover, for white noise the Gaussian Boltzmann equilibrium distribution in the harmonic potential implies a  spreading temperature $T_s$ which can  be generalized  in a nonequilibrium steady state as
\begin{equation}
\label{eq:(spreading_temperature)}
T_s = k a^2 / k_B .
\end{equation}
\noindent
Third a kinetic temperature can be defined in the steady state via
\begin{equation}
\label{eq:(kinetic_temperature)}
\Tkin= 2\Wkin(t\to\infty)/k_B .
\end{equation}
\noindent
In Figure~\ref{fig:(7)}b we compare these three temperatures $T$, $T_s$ and $\Tkin$ for white and persistent noise as a function of $\lambda$. For white noise and vanishing $\lambda$ we are in equilibrium and the three temperatures coincide. Moreover they remain practically identical for increasing $\lambda$. However, for active noise (when $T= \gamma v_0^2 \tau_p/k_B$) and for finite jerk, they differ in general and clearly $T_s$ and $\Tkin$ diverge at the localization-delocalization transition. For the parameters chosen in  Figure~\ref{fig:(7)}b active noise has a lower temperature than that for passive noise which is reflecting the same trend found for a completely overdamped oscillator.

We finally remark that the distribution of $\ddot x$, $\dot x$, $x$ and $u$ in the steady state is a  multivariate Gaussian with zero mean as the full stochastic process is a superposition of Gaussians.
 
\begin{table}[htbp]
\centering
\begin{adjustbox}{max width=\linewidth}
\begin{tabular}{|c | cc | cc | cc |}

    \hline 
    & \multicolumn{2}{c}{\textbf{active}} 
    & \multicolumn{2}{c}{\textbf{persistent}} 
    & \multicolumn{2}{c|}{\textbf{passive}} \\
    \hline
    \textbf{pure jerky} & 4 & 3 & 4 & 4 & 3 & 3 \\
    \hline
    \textbf{inertial jerky} & 4 & 1 & 4 & 2 & 3 & 1 \\
    \hline
    \textbf{full model} & 4 & 0 & 4 & 0 & 3 & 0 \\
    \hline
\end{tabular}
\end{adjustbox}
\caption{Summary of the different dynamical exponents governing the mean kinetic energy $\Wkin(t)$ for short (left) and long (right) times for various special cases and active, persistent and passive noise.}
\label{tab:(III)}
\end{table}

\section{Conclusions}
A new kind of active particles was proposed which are influenced by and susceptible to jerk. This implies a sensitive reaction on acceleration changes which speeds up the dynamical spreading considerably. The jerk-dependence breaks the basic second Newtonian law as the force now depends on acceleration  and therefore also the effective inertial mass can become negative. 
A simple but basic model to describe a jerky active particle was put forward and solved analytically for the particle mean-square displacement. It is dominated by high dynamical exponents characterizing an extremely high dynamical spreading which is anomalous and superballistic, typically not known from active matter. Such active jerks can be realized in experiments (both in macroscopic and colloidal set-ups) and belong to the large class of feedback controlled active particles \cite{Wendehenne,feedback} 
that have some kind of ``intelligence'' \cite{Liebchen,Goh1,Goh2}, ``delay'' \cite{delay1,delay2,Toepfer,Sussman,Kopp,Dago,Chatterjee,Venstra}, ``memory'' \cite{active_memory1,active_memory2,Netz,Grosberg,Mungan} or use ``information'' to decide about their future \cite{Cichos,Chen_information,quorum_sensing}. These smart active particles are more complex than standard active particle that just self-propel with a constant speed \cite{Ramaswamy}.

We have only discussed the most elementary active particle model, governed by the active Ornstein-Uhlenbeck process. There are many options for future research. Subsequent work should address more general structural and dynamical correlations beyond the MSD and MSJ within the same model based on the
 full Gaussian distribution functions of $u$, $x$, $\dot x$, and $\ddot x$. 

It would also be interesting to generalize jerky dynamics to two dimensions and to more general active Brownian particle models \cite{parental}. Although it is in principal straight-forward from the present work how to proceed along this direction, analytical solutions for a single particle will be more difficult or even lacking. When realizing a jerky active robot-like particle in a macroscopic experiment, Coulomb friction will play a considerable role \cite{deGennes,Hayakawa,deGennes2,PRE_Coulomb,Menzel,Coulombfriction} and one should generalize the basic model to include this kind of nonlinear friction as well. Moreover even higher order time derivatives and and different nonlinear feedback-couplings can be considered where our analysis should be applicable in principle as well. Finally also collections of many interacting jerky active particles can be explored. The effect of jerk dynamics on motility-induced phase separation \cite{Tailleur_review}, flocking \cite{Vicsek,flocking} and active turbulence \cite{Wensink} is still open.

\section*{acknowledgements}
I thank Maxim Root, Remi Goerlich, Lorenzo Caprini, Kristian S. Olsen,  Stephy Jose and Margaret Rosenberg for help and discussions. Funding within the German Research Foundation (DFG) within project LO 418/29-1 is gratefully acknowledged.

\appendix

\section{}
\label{App:B}
Here we give the explicit solution of the zeros $\omega_1$, $\omega_2$, $\omega_3$ for the characteristic cubic polynomial \eqref{eq:(CC)} $ \omega^3 +i \frac{m}{\lambda} \omega^2 -  \frac{\gamma}{\lambda} \omega -i \frac {k}{\lambda}$ according to Cardano’s formula \cite{Cardan}: 
\begin{equation}
    \label{eq:(solution)}
    \omega_j = iy_j -im/3\lambda \quad(j=1,2,3).
\end{equation}
Here $y_1 = u+v$, $y_2= \epsilon_+ u + \epsilon_- v$, $y_3= \epsilon_- u + \epsilon_+ v$ with $\epsilon_\pm = -1/2 \pm i\sqrt{3}/2$,
$u=\sqrt[3] { -q/2 + \sqrt{D}}$ and $v=\sqrt[3] { -q/2 - \sqrt{D}}$, $q=2(m/\lambda)^3/27 - m\gamma/3\lambda^2 + k/\lambda$ and the discriminant 
$D=(p/3)^3 + (q/2)^2$ where $p =  (\gamma/\lambda) - (m/\lambda)^2/3$. Here the notation $\sqrt[3] {-| \rho|^3} = - | \rho|$ for any real $\rho$ is taken. If the discriminant $D$ is positive there are non-vanishing real parts in two of the three $\omega_j$'s. For $D<0$ all three $\omega_j$'s have vanishing real parts. The condition for all three imaginary parts of the $\omega_j$‘s to be negative then translates into the following three conditions which have to be fulfilled simultaneously: 
\begin{equation}
    \label{eq:(conditions2)}
    Re(y_j)-m/3\lambda < 0 
\end{equation}
for $j=1,2,3$ where $Re$ denotes the real part of a complex number.

For the special case of vanishing friction $\gamma =0$, it can directly been concluded by inspection of the explicit solutions for the three $\omega_j$‘s that at least one of those has a positive imaginary part.
Complementarily, for $m=0$  and $\lambda\not=0$, $\gamma>0$, again at least one of the three $\omega_j$'s has a positive imaginary part. This can be quickly verified by the following: consider the equation $\omega^3 +i \frac{m}{\lambda} \omega^2 -  \frac{\gamma}{\lambda} \omega -i \frac {k}{\lambda} = (\ \omega-\omega_1)(\omega - \omega_2)(\omega-\omega_3)$. In order to produce a vanishing quadratic term in $\omega$ we immediately get the condition $\omega_1 + \omega_2 + \omega_3 = 0$ which implies $Im(\omega_1 + \omega_2 + \omega_3) = 0$. Excluding three purely oscillating solutions, at least one of these imaginary parts of the  $\omega_j$‘s    is non-zero. To fulfill the latter condition at least one other imaginary part has to have opposite sign. Consequently at least one of the three $\omega_j$'s must have a positive imaginary part.



%

\end{document}